\documentclass[prl,twocolumn,showpacs,preprintnumbers,amsmath,amssymb]{revtex4}
\usepackage{graphicx}
\usepackage{amssymb}

\begin{document}
\title{Microscopic theory of the 
proximity effect in superconductor-graphene nanostructures}
\author{P. Burset, A. Levy Yeyati and A. Mart\'{\i}n-Rodero}
\affiliation{Departamento de F\'{i}sica Te\'{o}rica de la Materia
Condensada C-V, Universidad Aut\'{o}noma de Madrid, E-28049
Madrid, Spain}

\begin{abstract}
We present a theoretical analysis of the proximity effect at a 
graphene-superconductor interface. We use a tight-binding model
for the electronic states in this system which allows  
to describe the interface at the microscopic level. Two different 
interface models are proposed: one in which the superconductor 
induces a finite pairing in the graphene regions underneath, thus 
maintaining the honeycomb structure at the interface and one that 
assumes that the graphene layer is directly coupled to a bulk 
superconducting electrode. We show that properties like the
Andreev reflection probability and its channel decomposition
depend critically on the model used to describe the interface.
We also study the proximity effect on the local density of 
states on the graphene. For finite layers we analyze the induced 
{\it minigap} and how it is reduced when the length 
of the layer increases. Results for the local density
of states profiles for finite and semi-infinite layers are presented.
\end{abstract}

\maketitle

\section{Introduction}
The possibility to isolate and perform direct transport measurements  
on few or even single graphite layers \cite{novoselov04}
has triggered a large activity
in the condensed matter community. The case of a single
layer of carbon atoms, known as graphene, is of particular interest
because of its unique electronic structure which, under certain 
conditions corresponds to massless Dirac fermions confined in two 
dimensions \cite{massless}.

On the other hand, the coupling to a superconductor provides an
interesting way to test the electronic properties of
graphene. 
In a recent work by Beenakker \cite{beenakker06} it was shown that 
for an ideal interface between a superconductor and graphene
an unusual type of Andreev reflection, in which the hole is specularly 
reflected, appears. Several other effects involving graphene and 
superconductors like Josephson transport \cite{titov06,zareyan07},
re-entrance effect \cite{ossipov07}, and quasiparticle transport mediated
by multiple Andreev processes \cite{cuevas07} have been theoretically
analyzed.

In addition to its effect on the transport properties, the coupling
to a superconductor also should produce a change in the electronic 
spectral properties and the induction of pairing correlations due to
the proximity effect.  
The recent experimental achievement of good contact between superconducting
electrodes and graphene layers \cite{heersche07}
open the possibility to explore the
proximity effect on these systems with great detail. 
Furthermore, experiments were the proximity effect on graphene could be
explored even with atomic scale resolution using STM are underway
\cite{saclay}. At present only results for the total density of
states in superconductor-graphene-superconductor structures
have been presented \cite{titov07}.

The present work is aimed to study in detail the interface between the 
superconductor and the graphene sheet. To this end we shall describe
the electronic structure of graphene at the level of
the tight-binding approximation. This description allows us to 
analyze the superconductor-graphene interface more microscopically as
compared to a description where the continuous limit leading to an
effective Dirac-Bogoliubov-De Gennes equation is taken from the 
start \cite{schomerus07}.
In the continuous description it is usually assumed that the presence of the
interface do not couple different valleys of the graphene band structure,
which could not be the case in an actual experimental situation.
Moreover, when the study is focused on finite size graphene sheets, 
a strong dependence on the geometry of the edges appears. Thus, 
different symmetry directions will have distinct behavior \cite{brey06}. For 
{\it zigzag} edges zero-energy surface states appear \cite{brey06} which 
could hide the effects of the coupling to a superconductor.

In this work we will concentrate on interfaces defined along
an {\it armchair} edge. We propose two different models 
for this interface: the first one assumes that graphene is coupled directly 
to a bulk superconducting electrode which does not maintain the honeycomb 
structure of the graphene sheet; the second model studies the possibility 
that one superconducting electrode on top of the graphene sheet induces a 
finite pairing amplitude and shifts the Fermi level of the graphene sheet 
far away from the Dirac point. As we discuss below, the two 
models lead to different behavior of the Andreev reflection probability
as a function of energy, wave vector and doping level. We further analyze
several aspects of the spectral properties of the graphene layer within 
the two models both for the finite and the semi-infinite case.

The rest of the paper is organized as follows: 
in Sec. II we introduce the tight-binding model for a graphene layer and we 
show the analytic expressions for the Green functions for a semi-infinite and 
a finite layer. In Sec. III the two different models for the interface 
with a superconductor are defined and a general expression for the 
self-energy, which provides the basis for the 
calculations of the following sections, is obtained. In Sec. IV we  
study the model dependence of the Andreev reflection processes. 
We also study, in Sec. V, the influence of the different interface 
models on the local density of states of a finite  
graphene layer coupled to a superconductor, analyzing in particular the
minigap which is induced in the case of metallic layers.
Results for the spatially resolved density of states for a semi-infinite 
graphene layer are presented in Sec. VI. The paper is closed with some
concluding remarks.

\section{Description of isolated graphene layers}

\begin{figure}
\includegraphics[scale=0.35,angle=-90]{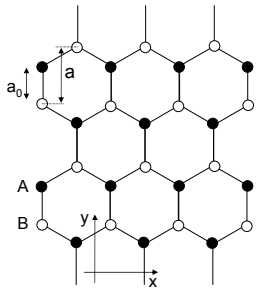}
\caption{The honeycomb structure of a graphene sheet is formed combining 
two triangular sublattices, denoted A (dots) and B (open dots). The unit cell 
on each horizontal line includes one atom from each sublattice. 
The axis selection used in this work is indicated.}
\label{figure1}
\end{figure}

For the description of the electronic states in a defect free graphene 
layer we shall adopt the tight-binding approximation, i.e. we use 
a model Hamiltonian of the type $\hat{H} 
= t_g \sum_{<ij>,\sigma} \hat{c}^{\dagger}_{i\sigma} \hat{c}_{j\sigma}
+ \epsilon \sum_{i\sigma} \hat{c}^{\dagger}_{i\sigma}\hat{c}_{i\sigma}$,
where $t_g$ denote the hopping element between nearest neighbors carbon
atoms on the hexagonal lattice and $\epsilon$ is a uniform site energy 
level which allows to vary the level of doping ($\epsilon=0$ corresponds
to the undoped case). The dispersion relation for the translational
invariant case is given by $E(k,q) = \epsilon \pm 
t_g |1 + 2 e^{iqa}\cos{(ka/\sqrt{3})}|$, where $k$ and $q$ denote the
wavevector in the $x$ and $y$ direction respectively and $a$ is the 
lattice parameter defined as shown in Fig. \ref{figure1} 
(as can be seen $a=3a_0/2$, where $a_0$ is the
interatomic distance).
For the undoped case the Fermi surface collapse
into two nonequivalent points at the Brillouin zone corresponding to 
$(k,q)=(\pm 2\pi/\sqrt{3}a,0)$. The dispersion relation close
to these points can be linearized with a slope $t_ga$
which fixes the Fermi velocity, $v_F$. 

\subsection{Green functions for a semi-infinite armchair graphene layer}

An essential ingredient for describing the interface between a graphene layer 
and other material is a good description of the electronic Green functions at 
the edges of the layer. We concentrate here in the derivation of the edge
Green function for a semi-infinite graphene layer with armchair orientation. 
We assume that there is translational symmetry in the direction parallel to 
the edge $(y)$. 
The semi-infinite system can be decomposed
into lines of sites in the $y$ direction which are coupled by hopping elements
with the neighboring lines on the $x$ direction. The unit cell on each line
includes two sites corresponding to each hexagonal sublattice that are 
denoted by A and B (see Fig. \ref{figure1}). 
These sites are coupled by a hopping element $t_g$ within the unit cell.
Thus, the cell Hamiltonian is given by

\[ \hat{h} = \left(\begin{array}{cc} \epsilon & t_g \\ t_g & \epsilon 
\end{array} \right). \] 

The hopping elements between neighboring lines couple also sites of type A with
sites of type B but should include a phase factor $e^{\pm iqa}$ due to the
displacement of the cells in the $y$ direction. The hopping matrix in the
A-B space (both in the forward and in the backward direction) can be written as
$\hat{t}(q) = t_g \hat{U}(q)$, where

\[ \hat{U}(q) = \left(\begin{array}{cc} 0 & e^{iqa} \\ e^{-iqa} & 0 
\end{array} \right). \] 

The self-similarity of the semi-infinite system with one additional line of 
sites leads to the following implicit equation for the edge Green function

\[ \hat{g}(q,\omega) = \left[ \omega \hat{I} - \hat{h} - t_g^2 \hat{U}(q) 
\hat{g}(q,\omega) \hat{U}(q) \right]^{-1} . \]

Hereafter we implicitly assume that $\omega$ stands for $\omega \pm i \eta$
and that the limit $\eta \rightarrow 0$ is taken to obtain the retarded or
the advanced component respectively.
We can now define $\hat{\tilde{g}} = \hat{U} \hat{g}$ which satisfies the
simpler equation

\[ \hat{\tilde{g}} = \left[ \hat{X}(q,\omega) - t_g^2 \hat{\tilde{g}} 
\right]^{-1} ,  \]
where $\hat{X}(q,\omega) = (\omega\hat{I} - \hat{h}) \hat{U}(q)$. 

To obtain an explicit expression for $\hat{\tilde{g}}$ 
it is useful to perform a basis
rotation in order to diagonalize the matrix $\hat{X}$. The general form of this
rotation is $\hat{R} = \hat{R}_1 \hat{R}_2$ where $\hat{R}_1 = 
e^{iqa/2\hat{\sigma}_z}$, $\hat{\sigma}_z$ being the $z$-Pauli matrix acting
on the sublattice space and 

\[ \hat{R}_2 = \frac{1}{\sqrt{-2\sin{\alpha}}}
\left( \begin{array}{cc} e^{i\alpha/2} & e^{-i\alpha/2} 
\\ ie^{-i\alpha/2} & ie^{i\alpha/2} \end{array} \right), \] 
where $\cos{\alpha} = t_g\sin{qa}/(\omega-\epsilon)$. 
The eigenvalues of $\hat{X}$ are $x_{1,2} = -t_g \cos{qa} \pm 
\sqrt{(\omega-\epsilon)^2 - t_g^2 \sin^2{qa}}$.   
We thus get

\begin{widetext}
\begin{eqnarray}
\hat{g}(q,\omega) & = & \left( \begin{array}{cc} g & f \\ f^{\prime} & g 
\end{array} \right)  
= \frac{1}{t_g} 
\hat{U} \hat{R}_1 \hat{R}_2 \left( \begin{array}{cc} e^{i\phi_1} & 0 \\ 0 & 
e^{-i\phi_2}
\end{array} \right) \hat{R}^{-1}_2 \hat{R}^{\dagger}_1 \nonumber \\
& = & \frac{1}{2 t_g \sin{\alpha}} 
\left( \begin{array}{cc} e^{i\phi_1}-e^{-i\phi_2} & 
ie^{iqa} \left[e^{-i(\alpha-\phi_1)} - e^{i(\alpha-\phi_2)} \right]\\  
-ie^{-iqa} \left[e^{i(\alpha+\phi_1)} - e^{-i(\alpha+\phi_2)} \right]   
&  e^{i\phi_1}-e^{-i\phi_2} \end{array} \right) ,
\end{eqnarray}
\end{widetext}
where $\cos{\phi_{1,2}} = x_{1,2}/(2t_g)$. The eigenvalues $e^{i\phi_1}$ and $e^{-i\phi_2}$ have been chosen so that the resulting Green functions have 
the proper behavior when the frequency goes to infinity. 

\subsection{Finite graphene layer}

Starting from the results of the previous section one 
can obtain the Green functions of a finite graphene layer by introducing a
perturbation consisting in breaking the bond between the $N$-th line and 
its neighbors on the $N+1$ line. From Dyson's equation we obtain the
following set of coupled equations 

\begin{eqnarray}
\hat{g}_{n,n}^F  &=&  \hat{g}_{n,n} - \hat{g}_{n,N+1} \hat{t} \hat{g}_{N,n}^F 
\nonumber \\
\hat{g}_{N,n}^F  &=&  \hat{g}_{N,n} - \hat{g}_{N,N+1} \hat{t} \hat{g}_{N,n}^F ,
\end{eqnarray}
where the superindex $F$ stands for the finite system and the subindexes 
$i,j$ indicate the lines within the layer. On the other hand the elements 
$\hat{g}_{N,N+1}$ can be expressed as 
$\hat{g}_{N,N+1} = \hat{g}_{N,N}^F \hat{t} \hat{g}_{N+1,N+1}$.
We now use that $\hat{g}_{1,1} = \hat{g}$ and
$\hat{g}_{n,N} = \hat{g}_{N,n} = \left(\hat{g} \hat{t} \right)^{N-n} \hat{g}_{n,n}$,
where $\hat{g}$ corresponds to the surface
Green function for the semi-infinite system derived in the previous section.
Also we have $\hat{g}_{N+1,N+1} = \left[ \hat{g} - \hat{t} \hat{g}_{N,N}^F 
\hat{t} \right]^{-1}$ and $\hat{g}_{n,n} = \left[ \hat{I} - \left(\hat{g} \hat{t} \right)^{2n}\right] \left[ \hat{I} - \left(\hat{g} \hat{t} \right)^{2}\right]^{-1} \hat{g}$, which allows
to obtain 

\begin{eqnarray}
\hat{g}_{n,n}^F & = & \left[ \hat{I} - \left(\hat{g}\hat{t}\right)^{2}\right]^{-1} \left[ \hat{I} - \left(\hat{g}\hat{t}\right)^{2(N+1)}
\right]^{-1}  \left[ \hat{I} - \left(\hat{g}\hat{t}\right)^{2n}\right] 
\nonumber \\
 & & \times \left[ \hat{I} - \left(\hat{g}\hat{t}\right)^{2\left(N-n+1\right)}\right] \hat{g} \\ 
\hat{g}_{n,N}^F & = & \left[ \hat{I} - \left(\hat{g}\hat{t}\right)^{2(N+1)}
\right]^{-1} \left[ \hat{I} - \left(\hat{g}\hat{t}\right)^{2n}\right] 
\left(\hat{g}\hat{t}\right)^{N-n} \hat{g} \nonumber.
\end{eqnarray}

Making use of the rotation matrix defined in the previous section these 
quantities can be written in the following rather simple form

\begin{widetext}
\begin{eqnarray}
\hat{g}_{n,n}^F & = & 
\frac{1}{t_g} \hat{R} \left( \begin{array}{cc} 
\frac{\sin{n\phi_1}}{\sin{\phi_1}} \frac{\sin{\left(N-n+1\right)\phi_1}}{\sin{(N+1)\phi_1}} & 0 \\
0 & \frac{\sin{n\phi_2}}{\sin{\phi_2}} \frac{\sin{\left(N-n+1\right)\phi_2}}{\sin{(N+1)\phi_2}} \end{array} \right) \hat{R}^{-1} 
\hat{U}\nonumber\\ 
\hat{g}_{n,N}^F & = & 
\frac{1}{t_g} \hat{R} \left( \begin{array}{cc} 
\frac{\sin{n\phi_1}}{\sin{(N+1)\phi_1}} & 0 \\
0 & \frac{\sin{n\phi_2}}{\sin{(N+1)\phi_2}} \end{array} \right) \hat{R}^{-1} 
\hat{U}.
\end{eqnarray}
\end{widetext}

One can have the expression for the borders of the layer setting $n=1$ or $n=N$. Then, the eigenvalues of the Green functions become $\sin{N\phi_i}/\sin{\left(N+1\right)\phi_i}$ and $\sin{\phi_i}/\sin{\left(N+1\right)\phi_i}$, with $i=1,2$, for the $\hat{g}_{1,1}^F=\hat{g}_{N,N}^F$ and $\hat{g}_{1,N}^F=\hat{g}_{N,1}^F$ cases, respectively.  This expressions are equivalent to those for a
finite tight-binding chain \cite{vecino01}.

The poles of these Green functions determine the spectral properties of
the layer. These poles are fixed by the condition 
$\sin{\left(N+1 \right)\phi_{1,2}} = 0 $, which 
is satisfied by $\phi_{1,2} = m \pi / (N+1)$, where $m$ is an 
integer. One can associate this condition with the quantization of
the transverse momentum which is used in the continuous model for
describing armchair nanoribbons \cite{brey06}. 
At the charge neutrality condition the existence of zero energy
states requires $\phi_{1,2} = \pm 2\pi/3$, which can only be
satisfied for $N=3p+2$ (in a more compact notation for $N \bmod 3 = 2$).
Therefore the layers can be classified into metallic, for the $N \bmod 3 = 2$ 
case, and insulating for the other cases 
($N \bmod 3 = 0,1$). In the insulating cases the gap in the spectrum 
is $2 E_g$, where 
$E_g \simeq \pi \hbar v_F/3L$, $L=Na/\sqrt{3}$ being the length of the layer.
It should be noted that electron states in the metallic case are doubly 
degenerate, while the degeneracy is removed in the insulating cases
\cite{brey06}.

\section{Modeling the graphene-superconductor interface}

One of the aims of the present work is to analyze different ways to describe 
the interface between a graphene layer and a superconductor. In the recent 
literature it has been assumed that a superconducting electrode deposited 
on top of graphene induces a finite pairing amplitude and introduces a 
finite level of doping which shifts the Fermi level of the graphene layer 
far away from the Dirac point \cite{beenakker06,titov06}. 
This heavily doped superconducting graphene (HDSC) model provides a 
simple boundary condition for the effective Dirac-Bogoliubov-De Gennes
equations describing the interface.  

One can alternatively imagine that the graphene layer is coupled directly 
to a bulk superconducting electrode by means of a sharp interface 
which breaks the coherence between the two graphene sublattices.  
We shall refer to this case as the {\it bulk-BCS} model.
We wish to analyze the 
differences between the two models.

The presence of superconducting correlations requires to introduce the
Nambu space describing electron and hole propagation within the graphene layer. 
All Green functions acquire a $2\times2$ structure in Nambu space. For the
uncoupled graphene we have

\begin{equation}
\check{\hat{g}} = \left(\begin{array}{cc} \hat{g}_e & 0 \\ 0 & \hat{g}_h 
\end{array} \right) , 
\label{nambu}
\end{equation}
where $\hat{g}_e$ corresponds to the propagators obtained in the previous 
section and $\hat{g}_h$ is obtained from $\hat{g}_e$ by changing 
$\epsilon \rightarrow -\epsilon$ and $t_g \rightarrow -t_g$ 
(notice that we 
use the hat symbol to denote the sublattice space while 
the check symbol indicates the Nambu space).

The effect of the coupling with the superconducting electrode can be introduced
by means of a self-energy $\check{\hat{\Sigma}}$ which renormalizes the
uncoupled Green functions. Thus, the local Green function on the graphene
edge at the interface is determined by
$[\check{\hat{g}}^{-1} - \check{\hat{\Sigma}}]^{-1}$. 

In the case of the bulk-BCS model 
the self-energy is momentum independent and does not have a 
structure in the sublattice space; i.e.  

\begin{equation}
\check{\hat{\Sigma}} = \pi \rho_s t_c^2 \check{\tau}_z \check{g}_{BCS} 
\check{\tau}_z \otimes \hat{I},  
\label{bulk-BCS}
\end{equation}
where $\check{\tau}_z$ is the $z$ Pauli matrix in Nambu space and
$t_c^2$ is the mean square
hopping element between the graphene layer and the superconducting electrode.
This quantity controls the value of the parameter $\beta 
= t_c^2 \pi \rho_s/t_g$ 
which
characterizes the quality of the the interface, $\rho_s$ being the density of
states. On the other hand, 
$\check{g}_{BCS} = g_s \check{I} + f_s \check{\tau}_x$
stands for the dimensionless BCS Green function, i.e. 
$g_s = -\omega/\sqrt{\Delta^2 - \omega^2}$ and 
$f_s = \Delta/\sqrt{\Delta^2 - \omega^2}$. 

We would like now to derive the expression of the self-energy 
within the HDSC model.
We first notice that in the heavily doped limit one has $|\epsilon| >> 
|\omega|,t_g \sin{qa}$ for the relevant range of frequencies and $q$ values. 
Thus in $\hat{R}_2$ one has $\alpha \rightarrow \pi/2$. In the new basis the
system is equivalent to a tight-binding chain with site energies 
$\epsilon \pm t_g$ and local pairing fixed by $\Delta$. Although the exact
Green functions for this system is rather complicated, for low energies 
it can be approximated by

\begin{eqnarray}
\hat{R}^{-1}_2 \check{\hat{g}} \hat{R}_2 &\simeq&  
\left( \begin{array}{cc} \pi \rho_+ & 0 \\ 0 & \pi \rho_- \end{array} \right)
\otimes \check{g}_{BCS} \nonumber \\
&& + \frac{1}{2t_g^2} \left( \begin{array}{cc} 
\epsilon+t_g & 0 \\ 0 & \epsilon-t_g \end{array}
\right) \otimes \check{\tau}_z , 
\end{eqnarray}
where $\pi \rho_{\pm} = \sqrt{1 - \left(\frac{\epsilon\pm t_g}{2t_g}\right)^2}/t_g$.
By further taking the approximation $|\epsilon| << t_g$ and transforming back
into the site representation we obtain the following self-energy

\begin{equation}
\check{\hat{\Sigma}}/t_g \simeq 
\frac{\sqrt{3}}{2} \check{\tau}_z \check{g}_{BCS} \check{\tau}_z \otimes \hat{I} - 
\frac{1}{2} \check{\tau}_z \otimes \hat{\sigma}_x  . 
\label{HDSC}
\end{equation}

Notice that, in contrast to the first model, the self-energy in the
HDSC model does exhibit a structure in the sublattice space. However, 
it satisfies the condition $\mbox{det} \check{\hat{\Sigma}}/t_g = 1$.
This structure turns out to be of importance in connection to Andreev 
reflection as discussed in the next section. In the following we will 
use a general form of the model self-energies, which can be expressed 
as $\check{\hat{\Sigma}} = \beta \check{\tau}_z 
\check{g}_{BCS}
\check{\tau}_z \otimes \hat{I} - \gamma \check{\tau}_z \otimes \hat{\sigma}_x$. 
Thus, appropriate values for $\beta$ and $\gamma$ will correspond to the 
different models (i.e. $\gamma=0$ with arbitrary $\beta$ 
for the bulk-BCS model, while $\beta = \sqrt{3}/2$ and $\gamma=1/2$
in units of $t_g$, for the HDSC model).

\section{Andreev reflection at a graphene-superconductor interface}

The Andreev reflection is the basic mechanism for the conversion of a
quasiparticle current into a supercurrent at the interface between a normal
metal and a superconductor. In the case of a graphene-superconductor interface
like the one we have described in the previous section there are two channels
for the incident electrons with a given wave vector $q$ corresponding to the
states which diagonalize the $\hat{X}$ matrix. The reflected hole can be in 
either of these two channels. Our microscopic theory can thus
describe a more general situation than the idealized model for the 
interface used in \cite{beenakker06} which
assumes only one channel for the reflected hole for a given
wavector.

The Andreev reflection amplitudes can be expressed in terms of Green 
functions.
Generalizing previous works \cite{cuevas96,cuevas98} 
we can derive the expression

\begin{equation}
\hat{r}_A(q,\omega) = 2 i 
\hat{A}^{1/2}_e
\left\{\check{\hat{\Sigma}} 
\left[ \check{\hat{I}} - \check{\hat{g}} 
\check{\hat{\Sigma}}\right]^{-1}\right\}_{eh} 
\hat{A}^{1/2}_h , 
\label{coef-RA1}
\end{equation}
where $\hat{A}_{e,h}(q,\omega) = \left(\hat{g}_{e,h}(q,\omega) - 
\hat{g}_{e,h}^{\dagger}(q,\omega)\right)/2i$.
Using the general form of the model self-energies discussed in the previous section allows us to reduce the expression of $\hat{r}_A$ to 

\begin{eqnarray}
\hat{r}_A & = & 2 i 
\hat{A}^{1/2}_e 
\beta f_s \left[\hat{I} - \beta g_s 
\left(\hat{g}_e + \hat{g}_h \right)  - \gamma \left(\hat{g}_h \hat{\sigma}_x
- \hat{\sigma}_x \hat{g}_e\right)  \right. \nonumber \\
& & \left. - 
(\beta^2 + \gamma^2) \hat{g}_h \hat{g}_e\right]^{-1} \hat{A}^{1/2}_h .  
\label{coef-RA2}
\end{eqnarray}

This expression becomes particularly simple when 
$\epsilon=0$ because $g^e=g^h$, $f^e=-f^h$ and 
$f^{\prime e}=-f^{\prime h}$. So $\hat{r}_A$ is an scalar quantity given by

\begin{equation}
\hat{r}_A = \frac{4 \beta f_s \left(\mbox{Im}g^2 - |f-f^{\prime *}|^2\right)}
{1 - \mbox{Tr} \left[ \left(\beta g_s \hat{I} \mp \gamma \hat{\sigma}_x \right) \hat{g}_{e,h} \right] - 
(\beta^2+\gamma^2) \det{\hat{g}_{e,h}} }  \hat{I} ,   
\label{coef-RA3}
\end{equation}
where $\mbox{Tr} \left[ \left(\beta g_s \hat{I} \mp \gamma \hat{\sigma}_x \right) \hat{g}_{e,h} \right] 
= \left[2 \beta g_s g - \gamma (f+f^{\prime}) \right]$ 
and $\det{\hat{g}_{e,h}} = (g^2 - f f^{\prime})$.

\begin{figure}
\includegraphics[scale=0.35]{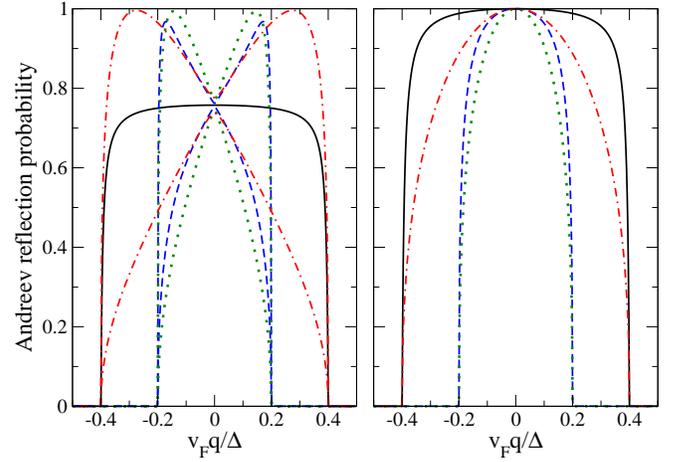}
\caption{Andreev reflection probability on the two eigenchannels 
as a function of the parallel momentum $q$ for fixed energy 
$\omega = 0.2 \Delta$. The left panel corresponds to the bulk-BCS model 
(with $\beta = 1$) and the right one to the HDSC model. The different 
curves correspond to different values of the doping level 
$\epsilon$: 0 (full lines), 0.1 (dashed lines), 
0.3 (dotted lines) and $0.4 \Delta$ (dashed-dotted lines).} 
\label{andreev-coef}
\end{figure}

In general doping conditions (i.e. when $\epsilon \ne 0$) $\hat{r}_A$
is not a scalar within the bulk-BCS model. The eigenvalues of
$\hat{r}_A \hat{r}^{\dagger}_A$ give the Andreev reflection probability
decomposed into two eigenchannels for each wavector $q$.
The evolution of these eigenvalues
for fixed $\omega$ and increasing $\epsilon$ as a function
of $q$ is shown in the left panel 
of Fig. \ref{andreev-coef}. 
Their maximum value is reached for $\beta=1$ and it never
exceeds $\sim 0.76$ at normal incidence. 
Within the HDSC model, however, $\hat{r}_A$
remains scalar for arbitrary doping and it always reaches the unitary
limit at $q=0$ (see right panel of Fig. \ref{andreev-coef}).

It is also interesting to analyze the physical character of the Andreev
reflection in the two models. The information on how the eigenchannels
of the uncoupled structure are connected by an Andreev process is 
contained in the matrix $\hat{R}_e \hat{r}_A \hat{R}^{-1}_h$.
As $\hat{r}_A$ is a scalar within the HDSC
model, the channel mixing is determined by $\hat{R}_e \hat{R}^{-1}_h$.
This is also the case for the bulk-BCS model at zero doping.
In this case we have $\alpha_h = \pi - \alpha_e$ and thus
electrons injected in one channel emerge as holes in the opposite one.
The momentum in the $y$ direction is conserved in this process and therefore
this type of reflection corresponds to 
what has been described in \cite{beenakker06} as 
{\it specular Andreev reflection}.
On the other hand for $\epsilon \ne 0$ and $\omega \rightarrow 0$ we have
$\alpha_h = \alpha_e$ which corresponds to the usual (retro) reflection where 
holes are reflected on the same channel as the incident electron. 
For intermediate doping situations both type of reflection would be
present although with a dominance of specular (retro) reflection for 
$\omega >\epsilon$ ($\omega < \epsilon$).

To complete the analysis of the present section we have computed the 
conductance per unit length due to Andreev processes, given by 
\cite{btk}

\begin{equation}
G_{AR} = \frac{2e^2}{h} \frac{1}{2\pi} 
\int_{-\frac{\pi}{a}}^{\frac{\pi}{a}}
dq \mbox{Tr} \left[ \hat{r}(q,eV) \hat{r}^{\dagger}(q,eV) \right] .
\end{equation}

\begin{figure}[hb!]
\includegraphics[scale=0.3]{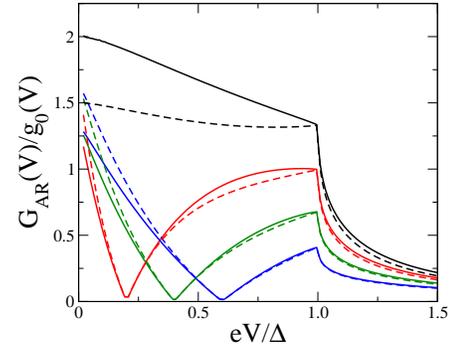}
\caption{Total conductance per unit length due to Andreev reflection 
$G_{AR}$ normalized to the conductance per unit length of a ballistic
graphene layer $g_0(V) = \frac{4e^2}{h} (eV+\epsilon)/(\pi \hbar v_F)$.
The results for the HDCS model (full lines) and for the bulk-BCS model
(dashed lines) are compared with increasing doping level 
$\epsilon = 0, 0.2, 0.4$ and $0.6 \Delta$.}  
\label{conductance}
\end{figure}

The results for $G_{AR}$ within the two models are shown in Fig. 
\ref{conductance}. As suggested in Ref. \cite{beenakker06} 
we normalize the result by the conductance per unit length
of a ballistic graphene sheet, which 
in the low energy limit is given by
$g_0(V) = \frac{4e^2}{h} (eV+\epsilon)/(\pi \hbar v_F)$.
As can be observed, at the charge neutrality point the HCSD 
model yields a maximum of ratio $G_{AR}/g_0=2$ at zero voltage,
which drops to $\sim 1.33$ for $\epsilon > 0$,
in agreement with the results of Ref. \cite{beenakker06}. This
ratio is of the order of $\sim 1.63$ in the bulk-BCS model 
with $\beta=1$ regardless of the doping level.
The qualitative behavior of the conductance 
with $\epsilon$ is similar in both models and agree
with the results of Ref. \cite{beenakker06}.

\section{Proximity effect on a finite graphene layer}

Using the previous results one can analyze the effect of the coupling
with the superconductor on the spectral properties 
of a graphene layer of finite size. Again, we will focus on the differences 
between the bulk-BCS and the HDSC models for the interface. 
>From the Dyson's equation it is straightforward to obtain the Green functions
at the edge of the layer (labeled as $1$) when the coupling to the 
superconductor is introduced on the opposite edge (labeled as $N$). 
Then, for an arbitrary line $n$ inside the layer we have

\begin{equation}
\check{\hat{G}}_{n,n} = \check{\hat{g}}_{n,n}^F + 
\check{\hat{g}}_{n,N}^F \check{\hat{\Sigma}} 
\left[\check{\hat{I}} - \check{\hat{g}}_{N,N}^F \check{\hat{\Sigma}} 
\right]^{-1} \check{\hat{g}}_{N,n}^F , 
\label{gnn-coupled-finite}
\end{equation}
where $\check{\hat{\Sigma}}$ stands for the general form of the 
self-energy introduced in Sec. III. We can further reduce this expression to

\begin{widetext}
\begin{equation}
\check{\hat{G}}_{n,n} = \check{\hat{g}}_{n,n}^F + \check{\hat{g}}_{n,N}^F 
\left( \begin{array}{cc}
\left[\beta g_s - \gamma \hat{\sigma}_x + \left(\beta^2 + \gamma^2 \right) \hat{g}^{Fh}_{N,N} \right] \hat{D}_e^{-1} &
- \beta f_s \hat{D}_h^{-1} \\
- \beta f_s \hat{D}_e^{-1} &
\left[ \beta g_s + \gamma \hat{\sigma}_x + \left(\beta^2 + \gamma^2 \right) \hat{g}^{Fe}_{N,N} \right] \hat{D}_h^{-1} 	
\end{array} \right) \check{\hat{g}}_{N,n}^F . 
\end{equation}
\end{widetext}

The quantities $\hat{D}_e$ and $\hat{D}_h$ have the following general form, expressed in the graphene subspace
\begin{eqnarray}
\hat{D}_e &=& \hat{I} - \beta g_s (\hat{g}^F_e + \hat{g}^F_h ) - \gamma (\hat{g}^F_e \hat{\sigma}_x - \hat{\sigma}_x \hat{g}^F_h ) \nonumber \\
&&- (\beta^2 + \gamma^2) \hat{g}^F_e \hat{g}^F_h
\nonumber \\
\hat{D}_h &=& \hat{I} - \beta g_s (\hat{g}^F_e + \hat{g}^F_h ) - \gamma (\hat{g}^F_h \hat{\sigma}_x - \hat{\sigma}_x \hat{g}^F_e ) \nonumber \\
&&- (\beta^2 + \gamma^2) \hat{g}^F_h \hat{g}^F_e \nonumber
\end{eqnarray}

As in the previous section, in the limit where $ \epsilon = 0 $ these 
two denominators become equal and simplify to an scalar, $D$, 
because $ \beta g_s (\hat{g}^F_e + \hat{g}^F_h) + \gamma (\hat{g}^F_e 
\hat{\sigma}_x - 
\hat{\sigma}_x \hat{g}^F_h) = \mbox{Tr} \left[\left(\beta g_s \hat{I} 
\mp \gamma \hat{\sigma}_x \right) \hat{g}^F_{e,h} \right] $, and 
$ \hat{g}^F_e \hat{g}^F_h = \det{\hat{g}^F} $. Then

\begin{equation}
D = 1 - \mbox{Tr} \left[ \left(\beta g_s \hat{I} \mp \gamma 
\hat{\sigma}_x \right) \hat{g}^F_{e,h} \right] - 
\left(\beta^2 + \gamma^2 \right) \det{\hat{g}^F_{e,h}}
\end{equation}

Using the expression of the finite layer Green functions given in section II
and the rotation given in section I one can easily show that

\begin{eqnarray}
\det \left[ \hat{g}^F_{e,h} \right] & = & - \frac {1} {t_g^2} \frac {\sin{\left(N \phi_1 \right)} \sin{\left(N \phi_2 \right)}}{\sin{\left[(N+1) \phi_1 \right]} \sin{\left[(N+1) \phi_2 \right]}}, \nonumber 
\end{eqnarray}
\begin{widetext}
\begin{eqnarray}
\mbox{Tr} \left[ \left(\beta g_s \hat{I} \mp \gamma \hat{\sigma}_x \right) \hat{g}^F_{e,h} \right] & = & \frac{1}{t_g \sin{\alpha}} \left\{ \frac {\sin{\left(N \phi_1 \right)} \sin{\left[(N+1) \phi_2 \right]} \left[\beta g_s - \gamma \sin{ \left( \alpha + q \right)} \right]} {\sin{\left[(N+1) \phi_1 \right]}\sin{\left[(N+1) \phi_2 \right]}} \right. \nonumber \\
& - & \left. \frac { \sin{\left(N \phi_2 \right)} \sin{\left[(N+1) \phi_1 \right]} \left[\beta g_s + \gamma \sin{\left(\alpha - q \right)} \right]} {\sin{\left[(N+1) \phi_1 \right]}\sin{\left[(N+1) \phi_2 \right]}} \right\}. \nonumber
\end{eqnarray}
\end{widetext}

The zeroes of $D$ determine the poles of the coupled system 
Green functions. From them one can thus analyze the distortion of the 
spectrum due to the superconducting proximity effect. For the charge 
neutrality case, this zeroes can be obtained from the expression

\begin{widetext}
\begin{eqnarray}
\frac { - \beta \omega^2}{\sqrt{\omega^2 - t_g^2 \sin^2{qa}} \sqrt{\Delta^2 - \omega^2}} & = & \frac {\sin{\left[(N+1) \phi_1 \right]}\sin{\left[(N+1) \phi_2 \right]} + (\beta^2 + \gamma^2) \sin{\left(N \phi_1 \right)} \sin{\left(N \phi_2 \right)}} {\sin{\left(N \phi_1 \right)} \sin{\left[(N+1) \phi_2 \right]} - \sin{\left(N \phi_2 \right)} \sin{\left[(N+1) \phi_1 \right]}} \nonumber \\
& + & \frac {\gamma} {\sin{\alpha}} \frac {\sin{(\alpha + q)} \sin{\left(N \phi_1 \right)} \sin{\left[(N+1) \phi_2 \right]} + \sin{(\alpha - q)} \sin{\left(N \phi_2 \right)} \sin{\left[(N+1) \phi_1 \right]}} {\sin{\left(N \phi_1 \right)} \sin{\left[(N+1) \phi_2 \right]} - \sin{\left(N \phi_2 \right)} \sin{\left[(N+1) \phi_1 \right]}} . \nonumber \\
\label{zeros}
\end{eqnarray}
\end{widetext}

\hspace{1cm}
\begin{figure}
\includegraphics[scale=0.32]{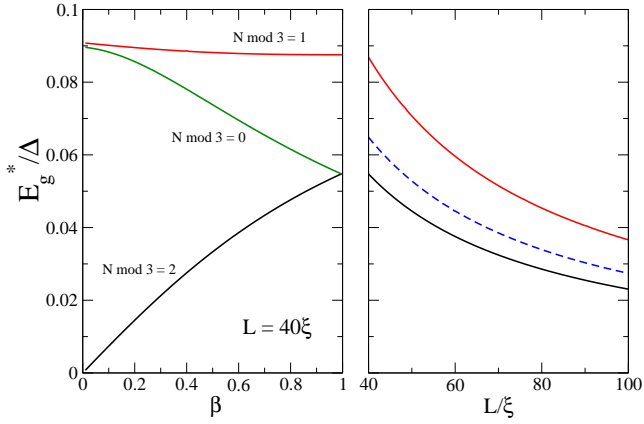}
\caption{Evolution of the lowest energy level $E^*_g$ of a finite
graphene layer coupled to a superconductor as a function of 
$\beta$ within the bulk-BCS model for $L = 40 \xi$
(left panel) and with the length of the layer $L$ within both the bulk-BCS (at $\beta = 1$) and the
HDSC models (right panel). 
In the case of the bulk-BCS model for arbitrary $\beta$ three 
different behaviors are found depending on the $N \bmod 3$ value. 
On the contrary, a universal
behavior of $E^*_g$ is found within the HDSC model regardless of the 
$N$ value (see text).}
\label{minigap}
\end{figure}

The spectrum corresponds to a series of subbands which disperse 
quadratically as a function of $q$ in the small $q$ limit. 
As in the uncoupled case the precise form of the dispersion relation 
depends on the value of $N \bmod 3$. In the case where the uncoupled
layer is metallic ($N\bmod 3= 2$) the coupling to the superconductor
induces a {\it minigap} in the lowest band. 
The existence of this minigap in the spectrum is similar to what is found 
for diffusive conductors and can be associated to the pseudo-diffusive
behavior of graphene at the charge neutrality point 
\cite{akhmerov07,prada07}.
For the cases $N\bmod 3=0,1$ the uncoupled layer is insulating and the
coupling of the superconductor just leads to a renormalization of the
gap in the spectrum. We shall denote by $E^*_g$ the lowest energy
level for all three cases.

The dependence $E^*_g$ as a function of $N$ and the
interface parameters $\beta$ and $\gamma$ can be obtained from Eq.
(\ref{zeros}) with $q=0$. For large $N$ this level decreases as 
$1/N$ with a prefactor which depends on $N\bmod 3$ and the interface 
parameters.
Fig. \ref{minigap} describes the behavior of $E^*_g$ both
in the bulk-BCS and in the HDSC models. The left panel shows the
lowest energy state within the bulk-BCS model as a function of the
interface transparency parameter ($\beta$) for fixed $N$. The three cases
$N\bmod 3 = 0,1,2$ are shown. As can be observed, in the metallic case
the minigap evolves from zero at $\beta=0$ to a maximum value 
at $\beta=1$. On the other hand, the two insulating cases exhibit
different behavior. While the starting value at $\beta=0$ is fixed by 
$E_g$ in both cases, in the case $N\bmod 3=0$
it decreases until it reaches the same value as the one of the $N\bmod 3=2$
case for $\beta=1$. On the other hand $E^*_g$ remains 
approximately constant
for $N\bmod 3=1$. This behavior indicates that the proximity effect is
almost negligible in this case.  The right panel shows the behavior
of the lowest energy state as a function of $N$ both in the bulk-BCS model
with $\beta=1$ and in the HDSC. The results are universal (i.e. independent
of the ratio $\Delta/t_g$) when plotted as a function of $L/\xi$, where
$\xi= \hbar v_F/\pi \Delta$ is the superconducting coherence length.
It is interesting to note
that while in the bulk-BCS model two limiting $1/L$ curves, corresponding
to $N\bmod 3=0,2$ and $N\bmod 3=1$, appear,
in the HDSC model $E^*_g$ lay on the same $1/L$ curve
regardless of $N$ (dashed line in Fig. \ref{minigap}).

\begin{figure}
\includegraphics[scale=0.33,angle=90]{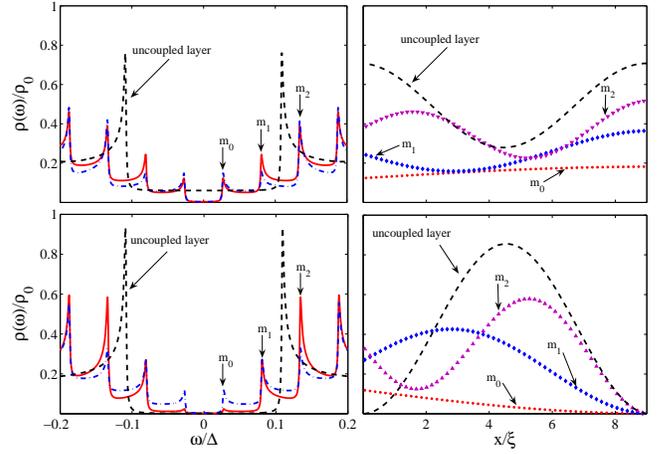}
\caption{Spatial variation of the LDOS on a finite layer ($L=9\xi$) in the 
HDSC model. The plots on the top panels correspond to the lines inside the 
layer with $n \bmod 3 = 1,2$ and those on lower ones correspond to  
$n \bmod 3 = 0$. The uncoupled layer 
has a metallic behavior ($N \bmod 3 =2$). The LDOS for this case at the
edges of the layer (top left panel) and at the center of the layer
(lower left panel) is plotted with dashed lines. 
When coupled to the superconductor, the LDOS is modified by the appearance of a 
minigap (denoted as $m_0$ in the pictures) and with the breaking of the 
uncoupled bands into a pair of subbands ($m_1$ and $m_2$ in the pictures). 
The evolution of the LDOS along the layer is shown in the right panels.
The results are normalized to the LDOS of a bulk graphene layer
with zero doping at $\omega=\Delta$, denoted by $\rho_0$.}
\label{ldos-finita}
\end{figure}

\subsection{Local density of states}
 
We define the electronic local density of states (LDOS) on a line 
$n$ within the graphene layer as

\begin{equation}
\rho_n(\omega) = \frac{a}{(2\pi)^2} \int_{-\frac{\pi}{a}}^{\frac{\pi}{a}} 
dq \mbox{Tr} \mbox{Im} \left[\hat{G}_{n,n}(q,\omega)\right],
\end{equation}
which has been normalized to one electron per site and spin. 
The LDOS thus defined is measured in units of $a/\hbar v_F$. 
However, to study the proximity effect it is more convenient to
normalize the LDOS with the density of a bulk graphene layer
with zero doping at $\omega=\Delta$, $\rho_0$, which 
for $\Delta \ll \hbar v_F/a$ is given by 
$\Delta/2\pi (a/\hbar v_F)^2$. The results thus obtained
do not depend on the choice of the ratio $\Delta/t_g$ used
in our tight-binding calculations.

The LDOS on a metallic layer ($N\bmod 3 =2$)
is shown in Fig. \ref{ldos-finita}. The results for the coupled case within
the HDCS model are compared with the results for the uncoupled case. 
It is typically found that the number of singularities in the LDOS
(associated with the number of subbands) in a given energy interval 
is doubled as compared to the uncoupled case. This effect is due to
the breaking of the double degeneracy of the bands due to the coupling
with the superconductor. 
The LDOS also exhibits an oscillatory behavior with the position on the layer.
This behavior reflects the properties of the electronic wave functions and,
as in the case of isolated nanoribbons \cite{brey06},
is distinct for lines with $n\bmod 3 =1,2$ and lines with
$n\bmod 3=0$. We thus illustrate these cases separately on the top and
on the lower panels of Fig. \ref{ldos-finita}. 

The right panels of Fig. \ref{ldos-finita} show the evolution along
the layer of the LDOS close to the singularities. We illustrate this evolution
at three different energies corresponding to the lowest first singularities, 
indicated by $m=0,1$ and 2 in Fig. \ref{ldos-finita}. For reference we
also show the spatial variation of the LDOS close to the first singularity
for the uncoupled case. In this case the LDOS reaches a maximum value at the
edges of the layer and a minimum in the middle for lines $n \bmod 3 = 1,2$, 
while the opposite behavior is found for $n \bmod 3 = 0$. 
In the coupled case one can
still identify the singularities with the oscillation pattern in the LDOS
but it does no longer reach an extreme value at the edge of the layer
in contact with the superconductor.

As final remark we note that in the case of insulating nanoribbons
the coupling to the superconductor just induces a shift  
singularities in the spectrum but do not change their number. This is due
to the nondegenerate character of the bands of the uncoupled layer.

\section{Proximity effect on a semi-infinite graphene layer}

The results of the previous section can be extended to analyze the spectral 
properties of a semi-infinite graphene layer coupled 
to a superconductor. The local Green function on a line $n$ is given by

 \[ \check{\hat{g}}_{n,n}  =  \left[ \check{\hat{G}}^{-1}_{n,n} - t_g^2 
\check{\hat{g}} \right]^{-1}, \]
where $\check{\hat{G}}_{n,n}$ is the Green function for a finite 
graphene layer 
coupled to a semi-infinite superconducting layer obtained in the previous 
section and $\check{\hat{g}}$ is the Green function for the edge of the 
semi-infinite layer.

\begin{figure}
\includegraphics[scale=0.5]{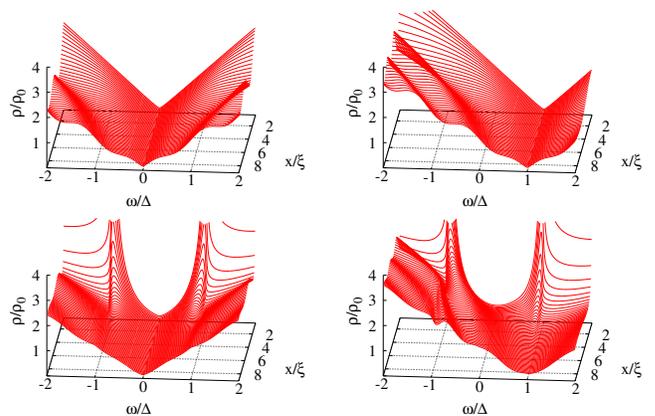}
\caption{LDOS on lines of type $n \bmod 3 = 1$ for a semi-infinite
graphene layer coupled to a superconductor within the bulk-BCS model
(lower panels). The upper panels show the corresponding results for
the uncoupled case. The plots on the left correspond to the undoped
case while those on the right correspond to $\epsilon=\Delta$.}
\label{dos-semiinfinite}
\end{figure}

\begin{figure}
\includegraphics[scale=0.3]{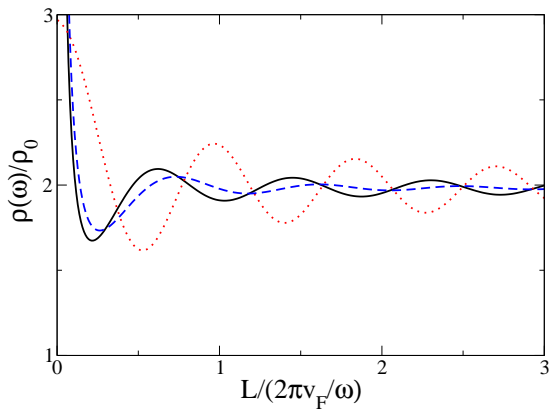}
\caption{Oscillation pattern on the LDOS of a semi-infinite
graphene layer at zero doping within the bulk-BCS model 
(full line), the HDCS model
(dashed line) and in the uncoupled case (dotted line).
The energy is fixed at $\omega= 2\Delta$.}
\label{ldos-oscillations}
\end{figure}

Fig. \ref{dos-semiinfinite} illustrates the spatial variation of the LDOS
on a semi-infinite graphene layer and the effect of varying the doping level
within the bulk-BCS model. For reference we show the
LDOS for the uncoupled case on the upper panels. As can be observed, 
the uncoupled LDOS  exhibits long wavelength oscillations
on the $\sim \hbar v_F/|\omega|$ scale on top of the characteristic 
V shape behavior. These oscillations are a surface effect which 
decreases in amplitude inside the layer as shown in Fig. 
\ref{ldos-oscillations}, where the LDOS profile at $\omega=2\Delta$
is plotted on a larger scale. A similar effect has been shown to occur
in the case of nanoribbons with zigzag edges \cite{tkachov07}.

The superconducting proximity effect is manifested by the appearance of
sharp peaks in the LDOS for energies $|\omega| \sim \Delta$ 
(lower panels on Fig. \ref{dos-semiinfinite}).
These peaks distort the V shape density of 
states, an effect which decays within a few times the coherence
length inside the layer. The small oscillations on the 
$\hbar v_F/|\omega|$ scale are reduced as compared to the uncoupled 
case but are still observable within the bulk-BCS model (indicated
by the full line in Fig. \ref{ldos-oscillations}).

The overall behavior of the LDOS within the HDSC model is very similar, 
although in this last case the $\hbar v_F/|\omega|$ oscillations are
further suppressed (dashed line in Fig. \ref{ldos-oscillations}).

The right panels in Fig. \ref{dos-semiinfinite} illustrate the effect 
on the LDOS of a displacement from the charge neutrality condition 
by applying a gate potential (finite $\epsilon$).
It is observed that the V shape is essentially rigidly displaced   
while the peaks induced by 
the proximity effect remain fixed at $\omega \sim \pm \Delta$. 
On the other hand, the wavelength of the oscillation pattern is 
in this case set by $\hbar v_F/|\omega-\epsilon|$.

\section{conclusions}

We have presented a theoretical analysis of the proximity effect at a 
graphene-superconductor interface. For this study we have first derived
analytical expressions for the Green functions on an armchair edge of
a semi-infinite graphene layer and for a finite layer.

Two models for describing microscopically the coupling  
to a superconductor have been presented. In the first model a bulk 
superconducting electrode is connected directly to the armchair 
edge of a graphene 
sheet (bulk-BCS model). The honeycomb structure of graphene is broken 
and this is reflected in the different behavior of the Andreev 
reflection probability on the 
two eigenchannels of the graphene sheet as a function of the 
parallel momentum $q$.
Only for the case of zero doping both eigenchannels are equivalent. 
Within this model
one can study the effect of varying the normal transparency 
of the interface (through parameter $\beta$). The Andreev reflection 
probability at normal incidence 
never reaches unity within this model but has a maximum value of $\sim 0.76$. 
In the second model it is assumed that the superconducting electrode 
induces a finite order parameter on the graphene regions underneath.
This model thus maintains the graphene sublattice structure and, as we have 
shown, 
the Andreev reflection amplitude $\hat{r}_A(q,\omega)$ is a scalar quantity
for arbitrary doping and always reaches the unitary limit for normal incidence.

We have also studied the effect of the coupling to the superconductor on the
spectral properties of finite and semi-infinite graphene layers. For finite 
layers we have obtained a simple expression for the energy spectrum of the 
coupled system 
which can be easily evaluated numerically. We have shown that a metallic ribbon 
develops a minigap whose size decreases inversely with the length of the layer. 
This effect can be associated with the pseudo-diffusive behavior of graphene. 
The induced minigap is slightly smaller in the bulk-BCS model with 
$\beta=1$ than in the HDSC model.

For the semi-infinite case the proximity effect manifests in the appearance of  
peaks in the density of states for frequencies $|\omega| \sim \Delta$. These 
peaks decay rapidly inside the graphene sheet for distances a few times 
the superconducting coherence length $\xi$. On the other hand, the LDOS 
keeps its characteristic V shape for zero doping for frequencies 
$|\omega| < \Delta$. We expect that these findings can be useful to 
analyze future STM experiments on graphene sheets with superconducting
electrodes.

\section{acknowledgments}
This work has been financed by the Spanish CYCIT (contract FIS2005-06255). 
Discussions with J.C. Cuevas, L. Brey and H. Le Sueur are acknowledged.

\end{document}